\documentclass[sigconf, screen]{acmart}
\usepackage{hyperref}
\usepackage{algorithmic}
\usepackage{graphicx}
\usepackage{textcomp}
\usepackage{xcolor}
\usepackage{fancyhdr}
\usepackage{etoolbox}
\usepackage{xspace}
\usepackage{booktabs}
\usepackage{comment}
\usepackage{makecell}
\usepackage{listings}

\usepackage{tikz}
\newcommand*\circled[1]{\tikz[baseline=(char.base)]{
            \node[shape=circle,draw,inner sep=.75pt] (char) {#1};}}

\newcommand{\SystemName}{\textit{ShEF}\xspace}

\newtoggle{showmarks}

\iftoggle{showmarks}{
  \newcommand\markyz[1]{\textcolor{violet}{[~MARK:~#1~]}}
  \newcommand\mingyu[1]{\textcolor{blue}{[~MINGYU:~#1~]}}
  \newcommand\christos[1]{\textcolor{orange}{[~christos:~#1~]}}
  \newcommand\blue[1]{\textcolor{blue}{#1}}
}{
  \newcommand\markyz[1]{\unskip}
  \newcommand\mingyu[1]{\unskip}
   \newcommand\christos[1]{\unskip}
  \newcommand{\blue}[1]{#1}
}

\newcommand{\revised}[1]{\textcolor{black}{#1}}

\AtBeginDocument{%
  \providecommand\BibTeX{{%
    \normalfont B\kern-0.5em{\scshape i\kern-0.25em b}\kern-0.8em\TeX}}}

\setcopyright{acmlicensed}
\acmPrice{15.00}
\acmDOI{10.1145/3503222.3507733}
\acmYear{2022}
\copyrightyear{2022}
\acmSubmissionID{asplos22main-p443-p}
\acmISBN{978-1-4503-9205-1/22/02}
\acmConference[ASPLOS '22]{Proceedings of the 27th ACM International Conference on Architectural Support for Programming Languages and Operating Systems}{February 28 -- March 4, 2022}{Lausanne, Switzerland}
\acmBooktitle{Proceedings of the 27th ACM International Conference on Architectural Support for Programming Languages and Operating Systems (ASPLOS '22), February 28 -- March 4, 2022, Lausanne, Switzerland}

\begin{document}

\title{\SystemName: Shielded Enclaves for Cloud FPGAs}

\author{Mark Zhao}
\email{myzhao@cs.stanford.edu}
\affiliation{
  \institution{Stanford University}
  \city{Stanford}
  \state{California}
  \country{USA}
}

\author{Mingyu Gao}
\email{gaomy@tsinghua.edu.cn}
\affiliation{
  \institution{Tsinghua University}
  \city{Beijing}
  \country{China}
}

\author{Christos Kozyrakis}
\email{christos@cs.stanford.edu}
\affiliation{
  \institution{Stanford University}
  \city{Stanford}
  \state{California}
  \country{USA}
}

\renewcommand{\shortauthors}{Mark Zhao, Mingyu Gao, and Christos Kozyrakis}

\begin{abstract}
FPGAs are now used in public clouds to
accelerate a wide range of applications, including many that operate
on sensitive data such as financial and medical records. We present
\SystemName, a trusted execution environment (TEE) for cloud-based
reconfigurable accelerators. \SystemName is independent from CPU-based TEEs and allows secure execution under a threat model where the adversary can control all software running on
the CPU connected to the FPGA, has physical access to the FPGA, and
can compromise the FPGA interface logic of the cloud provider. \SystemName
provides a \textit{secure boot and remote attestation} process that
relies solely on existing FPGA mechanisms for root of trust. It also includes
a \textit{Shield} component that provides secure access to data
while the accelerator is in use. The Shield is highly customizable and extensible, allowing users to craft a bespoke security solution that fits their accelerator's memory access patterns, bandwidth, and security requirements at minimum performance and area overheads.
We describe a prototype implementation of \SystemName
for existing cloud FPGAs, map \SystemName to a performant and secure storage application, and measure the performance benefits of customizable
security using five additional accelerators. 
\end{abstract}

\ccsdesc[500]{Computer systems organization~Reconfigurable computing}
\ccsdesc[500]{Security and privacy~Hardware security implementation}
\ccsdesc[500]{Computer systems organization~Cloud computing}

\keywords{trusted execution, enclaves, cloud computing, FPGAs, reconfigurable computing}

\maketitle

\section{Introduction}
\label{sec:intro}

Cloud computing is a double-edged sword. Cloud servers provide unmatched capabilities that are highly available, easily deployable, and widely scalable. This flexibility is critical for data-driven applications such as machine learning. However, the massive volume of data flowing through shared infrastructure presents new security and privacy concerns due to the corresponding growth in the trusted computing base (TCB) of cloud applications. When users process sensitive data on the cloud, they implicitly trust a multitude of entities including the developers and operators of the underlying hardware, OSes/VMMs, storage and database systems, and identity and access management services. They also trust the employees of the cloud service provider (CSP) who have physical access to the infrastructure. Recent data leaks demonstrate that a vulnerability in any layer of the stack can result in exposure of highly-sensitive information, e.g., the health and financial records for hundreds of millions of people~\cite{equifax, capital-one, jama-liu}.

These issues have led to the development of software security mechanisms to protect sensitive data in the cloud. Users can use purely cryptographic solutions such as homomorphic encryption (HE)~\cite{gentry-thesis} or integrate ad-hoc cryptographic schemes from a plethora of libraries. Unfortunately, HE is prohibitively expensive for most real-world applications~\cite{msr-ppa}.
Traditional cryptographic libraries, assuming correctness, still depend on a large TCB, including the CSP's multiple layers of software controlled by potentially malicious employees. %

CPU-based trusted execution environments (TEEs), such as Intel SGX~\cite{intel-sgx-explained} and ARM TrustZone~\cite{arm-trustzone} shrink the TCB.
TEEs provide hardware-based isolation for user code and data even given malicious privileged software and physical attacks.
However, hardware is inherently \textit{hardened}, presenting many security problems.
First, cryptography and cryptanalysis are constantly evolving; standards and best-practices constantly improve, especially as new computing paradigms such as quantum computing are introduced~\cite{post-quantum-crypto}.  
Second, applications use numerous compute and communication patterns, each demanding different levels of protection. 
For example, some applications may only require authenticated encryption for streaming data, while others read and write multiple times to a given address and thus need additional security such as Merkle Trees~\cite{suh-micro03}.
Finally, the recent vulnerabilities that directly compromise SGX highlight the difficulty of implementing bug-free security mechanisms in modern CPUs~\cite{meltdown, spectre, sgxpectre, spectre-returns, foreshadow, lvi, zombieload, ridl, fallout, plundervolt, microscope}.
\blue{For these reasons, it is desirable to have the flexibility to enhance security mechanisms post-manufacturing, a trait that hardened CPU security mechanisms do not provide.}

Moreover, CPU-based TEEs do not directly enable isolated execution on accelerators such as GPUs~\cite{hotchips-gpu}, FPGAs~\cite{hotchips-fpgaintel,hotchips-fpgaxilinx}, or TPUs~\cite{isca17-tpu}.
Slowing trends in process scaling are driving specialized hardware in order to achieve scalable performance~\cite{isscc14-horowitz}.
As a result, CSPs including Amazon~\cite{f1}, Microsoft~\cite{azure-fpga}, Huawei~\cite{huawei-fpga}, and Baidu~\cite{baidu-fpga}, are rapidly deploying remote FPGAs to meet increased computing demands, allowing users to deploy custom accelerators generated from application code~\cite{cad11-cong,pldi18-spatial,fpl19-hadjis,dnnweaver,iccad20-esp}. 

Supporting secure computing on remote FPGAs is thus essential for many emerging applications, such as medical, financial, machine learning (ML), and regulatory use-cases, that require both security and acceleration.
\blue{For example, FPGAs are at the forefront for accelerating genomics sequencing~\cite{asplos18-darwin, hpca19-wu}.
FPGAs are also extensively used for accelerating Deep Neural Networks (DNNs)~\cite{access19-shawahna}, which have demonstrated significant impact in medical diagnosis~\cite{nature-esteva}.
Recent work even directly call for accelerator TEEs, including FPGAs, as an essential component of distributed federated learning in the cloud~\cite{socc21-citadel} and GDPR-compliant storage systems~\cite{vldb21-sdp}.
These sensitive accelerators are seeing widespread adoption in industry, with Xilinx marketing FPGAs directly for ML-based diagnostics~\cite{versal-applications} and myriad companies offering AWS F1 accelerators for genomics~\cite{deneb, dragen, huxelerate}, financial analysis~\cite{f1-montecarlo}, and network security~\cite{valtix} applications on the AWS Marketplace.}

Unfortunately, recently-proposed TEEs for accelerators, including FPGAs, are either insecure against direct physical attacks~\cite{jang-asplos19, sp20-hetee}, require fundamental hardware changes~\cite{sp20-hetee, jang-asplos19, volos-osdi18}, only address isolated challenges such as attestation~\cite{eguro-fpl12, pfc, meetgo, ambassy}, or rely on external CPU TEEs~\cite{volos-osdi18, jang-asplos19, ambassy}. Moreover, they ignore the Shell logic~\cite{aws-fpga-dev-kit, osdi20-korolija, amorphos, trimberger-ivsw18}, a fundamental untrusted operating system for cloud FPGA logic.

We address these challenges with the {\it Shielded Enclaves for Cloud FPGAs} (\SystemName)
framework that brings together hardware-based \textit{bespoke security} and \textit{customizable acceleration} for FPGAs.
\SystemName applies the confidentiality, integrity, freshness, and isolation guarantees provided by TEEs~\cite{trustcom-tees} to custom FPGA accelerators, even in the presence of malicious software and hardware logic, or physical attacks. \SystemName targets current cloud FPGA deployments and commodity FPGA hardware. While \SystemName relies on CPUs for networking and data transfer, it is decoupled from and assumes no trust in CPU TEEs or other software running on CPUs. Hence, \SystemName minimizes the TCB and avoids the aforementioned issues with \emph{hardened} hardware.
\SystemName provides {\it customization} as a key feature, enabling users to adapt security mechanisms to match their accelerator's unique bandwidth requirements, memory access characteristics, and threat models. By provisioning only the right levels of protection, \SystemName allows users to address their threat model at minimum performance and area cost.

\SystemName consists of two main components.
The {\it \SystemName boot process} centers around a software security kernel that builds a chain of trust on top of currently-existing FPGA security mechanisms. %
Its primary purpose is to (\textit{a}) load the accelerator into a known and trusted state on the FPGA, (\textit{b}) attest the state to a remote verifier {\it chosen by the user}, and (\textit{c}) ensure that sensitive ports such as JTAG are secured during runtime.

Once the accelerator design is securely booted onto the FPGA, the {\it \SystemName Shield} communicates with host software and protects the accelerator's sensitive data through a highly customizable and extensible set of soft-logic engines.
Users can customize a rich set of parameters, such as encryption logic parallelism, optimizations for memory access patterns, cryptographic primitives, authentication block size, and key size over individual regions of memory.
For example, a DNN accelerator that reads in large blocks of weights can choose to use large encryption blocks to amortize integrity check overheads and \revised{forgo} expensive replay-attack countermeasures specifically for those weights.
Other accelerators that perform multiple small reads and writes (e.g., for graph processing) may conversely select smaller block sizes to prevent unnecessary data transfers and use heavyweight memory authentication techniques~\cite{hardware-memory-authentication}.

In summary, we make the following contributions:
\begin{itemize}
    \item \blue{We identify requirements on enabling TEEs on cloud FPGAs and gaps not addressed by prior work due to current cloud FPGA devices and environments.}
    \item \blue{We implement an end-to-end workflow on \textbf{current} FPGA devices that provides the first comprehensive and customizable TEE for cloud FPGAs.}
    \item \blue{We present protocols that enable important TEE building blocks, including secure boot and remote attestation, on top of our workflow.}
    \item \blue{We identify customizability as a key requirement for cloud FPGAs and provide a solution via the modular Shield to easily customize \SystemName to the diverse security and performance requirements of FPGA-based accelerators.}
\end{itemize}

\blue{We implemented the end-to-end \SystemName workflow on a representative FPGA, and we configured the Shield's parallelism and security for six accelerators with diverse performance requirements and access patterns, including a GDPR secure storage benchmark~\cite{vldb21-sdp} and DNNWeaver~\cite{dnnweaver}.
We demonstrated that \SystemName minimized overheads to 0-122\% with 3.1-11\% area on AWS F1 instances.
\SystemName is open-source\footnote{https://github.com/stanford-mast/ShEF}, allowing the community to review and build on its design.}
\section{Background and Motivation}
\label{sec:motivation}
\SystemName is motivated by the confluence of two important trends~\cite{turing-lecture}.
First, hardware security is becoming a first-class citizen, driven by the ongoing microarchitectural side-channel attacks to CPUs~\cite{meltdown, spectre, spectre-returns, zombieload, fallout} and SGX in particular~\cite{sgxpectre, plundervolt, foreshadow, ridl, lvi, microscope}.
Second, the end of Dennard scaling is galvanizing a shift towards energy-efficient domain-specific accelerators (DSAs).
It is unclear how to bridge the gap and provide secure computation for DSAs which have disparate security and performance requirements.

\SystemName realizes recent requests for FPGA TEEs in order to provide secure remote storage~\cite{vldb21-sdp}.
However, \SystemName goes further and makes the key insight to leverage the flexibility of cloud FPGAs to enable secure remote acceleration.
Numerous applications such as machine learning~\cite{fpl19-hadjis}, genomics~\cite{hpca19-wu}, multi-party computation~\cite{fpl20-wolfe}, and simulation~\cite{firesim} already utilize DSAs on cloud FPGAs. 
The goal of \SystemName is to provide security guarantees with the same level of flexibility and to create bespoke TEEs for cloud FPGAs that fit accelerators' specific needs.
TEEs, FPGA security, and cloud FPGAs are all well-studied domains.
Merging these domains, however, introduces the new challenges discussed in this section.

\subsection{Trusted Execution Environments (TEEs)} \label{sec:tees-background}
TEEs secure remote computation by providing an isolated environment for users' sensitive code on devices controlled by untrusted parties and running untrusted, privileged software.

While there exist numerous flavors of TEEs~\cite{phantom, sanctum_security16, splittingif_osdi06, oasis_ccs13, isox_micro14, hyperflow-ccs18, xom, suh-micro03, bastion, intel-sgx-explained, arm-trustzone,eurosys20-keystone}, there are necessary and sufficient building blocks that they all provide~\cite{trustcom-tees}.
TEEs are built on a chain of trust, starting with a \textbf{hardware-based root of trust} in the form of a private key stored securely on chip~\cite{trusted-pufs, trustcom-tees}.
A \textbf{secure boot} process extends trust by cryptographically measuring each component during boot, up to and including the secured application.
This integrity measurement is then cryptographically proven to a remote user of the secured application in a \textbf{remote attestation} process.
Once trust is established, confidential data is emplaced into the TEE from a \textbf{secure storage and I/O} channel.
The secure application processes the data, and the TEE ensures that any interaction with the rest of the system is secured via an \textbf{isolated execution} mechanism.

In the context of spatial computing platforms, such as FPGAs, we assert that a TEE must also be \textbf{customizable}.
Accelerators use a selection of I/O interfaces, exhibit a broad range of memory access and throughput characteristics, and require different levels of security mechanisms.
A static TEE that provides a general solution for all accelerators is doomed to either be over-instrumented and waste resources or not satisfy each accelerator's stringent throughput and security requirements.

\subsection{Conventional FPGA Security Mechanisms}\label{sec:fpga-security}
Xilinx~\cite{xilinx-applications} and Intel~\cite{intel-applications} FPGAs target mission-critical applications such as defense and networking, and thus adopt a similar threat model and security mechanisms~\cite{intel-sdm, xilinx-ultrascale-security}.
Namely, a single bitstream developer must have physical and secure access to the FPGA before deploying the device into an untrusted environment.
The security mechanisms' goal is to ensure that (a) only developer-signed bitstreams can be loaded, (b) bitstreams are encrypted to prevent reverse-engineering, and (c) the FPGA can detect and respond to physical tampering.
These mechanisms are enabled in Intel and Xilinx FPGAs via a series of redundant, embedded processor modules  executing from BootROM and programmable firmware~\cite{intel-sdm, xilinx-ultrascale-security}.
We refer to these as the Security Processor Block (SPB) hereinafter.

The SPB has access to two pieces of information embedded in secure, on-chip, non-volatile storage: an AES key and the hash of a \emph{public} ECDSA (Intel) or RSA (Xilinx) key.
The AES key can be further encrypted via a physically-unclonable function (PUF), preventing the AES key from being compromised under physical attacks.
The device owner or IP developer is meant to securely embed the necessary keys offline prior to deployment.
Depending on whether the developer desires encryption and/or authentication, she can encrypt the bitstream with the AES key and/or sign it with the ECDSA/RSA private key.
The bitstream can then be securely decrypted (using the AES key) and authenticated (using the public key hash) by the SPB. %
Finally, the SPB actively monitors for any tampering.

\subsection{Remote FPGAs-as-a-Service}\label{sec:faas}
\begin{figure}[t]
  \centering
  \includegraphics[width=3.33in]{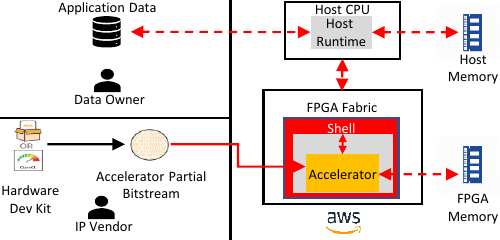}
  \caption{\small The AWS F1 development process and corresponding \SystemName threat model. The red arrows represent untrusted channels.}
  \label{fig:threat-model} 
\end{figure}

FPGAs are becoming an increasingly important component for major cloud service providers (CSPs)~\cite{f1, azure-fpga, huawei-fpga, baidu-fpga}.
The most prevalent example is AWS EC2 F1 instances, which provide 1, 2, or 8 exclusive Xilinx Virtex UltraScale+ VU9P FPGAs, each with 64GB of local DDR4 memory, tethered to a host CPU via a dedicated PCIe x16 link.

F1 instances are currently offered in a traditional Infrastructure-as-a-Service (IaaS) fashion as shown in Figure~\ref{fig:threat-model}.
IP vendors use the AWS FPGA Development Kit~\cite{aws-fpga-dev-kit} to create a custom accelerator using Xilinx's development tools such as RTL, OpenCL, or C/C++ High-level Synthesis (HLS).
Once the accelerator design is finalized, the developers compile it into a bitstream binary.
AWS leverages partial reconfiguration, which allows disjoint spatial regions of the FPGA fabric to be programmed via separate partial bitstreams.
An F1 instance is configured with two partial bitstreams: one belonging to the CSP which contains the \textit{Shell} logic, and one belonging to the user's accelerator design.
The Shell is analogous to an operating system as it provides the accelerator with virtualized peripherals and debugging features (virtual JTAG/LEDs), while protecting the physical FPGA from malicious logic~\cite{trimberger-ivsw18}.
The Shell is static logic and continuously runs on the FPGA.
At design time, developers connect their accelerator module's I/O ports to the standard Shell interfaces.
At deployment time, users leverage a command line interface to dynamically program their chosen partial bitstream onto the remaining reconfigurable region.
Once an accelerator is programmed, a runtime program on the host CPU initiates the accelerator and facilitates execution by transferring data between the CPU memory and FPGA device memory.

\subsection{Challenges for Secure and Customized Computing}\label{sec:challenges}
\SystemName aims to enable secure and customized computing on cloud FPGAs by providing key TEE building blocks. %
However, fundamental assumptions made by FPGA manufacturers and inherent differences between FPGA and CPU cloud offerings induce important challenges \blue{not addressed by prior work}.

\begin{figure*}[ht!]
\centering
\includegraphics[width=7in]{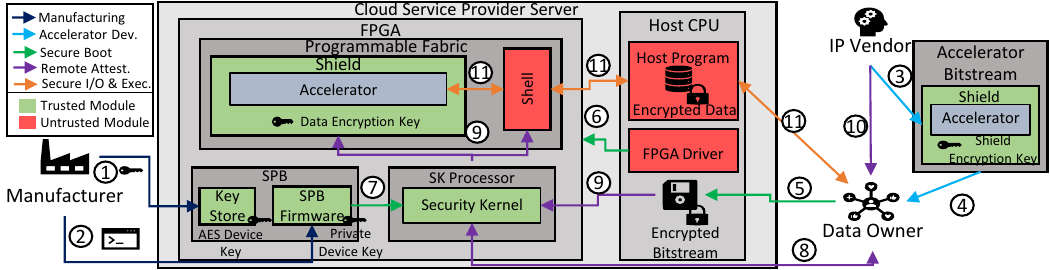}
\caption{\small \revised{Architecture of the \SystemName workflow with color-coded legend.}}
\label{fig:architecture}
\end{figure*}

\noindent\blue{\textbf{A lack of asymmetric keys.}
As described in Section~\ref{sec:fpga-security},  FPGA manufacturers assume a threat model for a single, trusted user whose AES key and public key hash are loaded onto the FPGA in a secure facility.
In contrast, cloud TEEs need to be utilized by multiple, mutually distrusting parties who never have physical access to the FPGA.
{\it Thus, an FPGA TEE must build a hardware root-of-trust and remote attestation protocol on top of the available AES key, as opposed to traditional private keys assumed by CPU and prior accelerator TEEs~\cite{trustcom-tees, intel-sgx-explained, volos-osdi18, jang-asplos19, sp20-hetee, nsdi20-telekine}}.}

\noindent\blue{\textbf{Presence of an untrusted Shell.}
Furthermore, many mechanisms CPUs use to enable TEE building blocks do not apply to the spatial architecture of FPGAs.
In CPUs, threads belonging to both secure and insecure applications are time-sliced on the processor. 
Thus, enclaves can directly access secure hardware using ISA extensions~\cite{intel-sgx-explained}, bypassing the untrusted OS.
For example, the SGX ISA extensions allow users to directly access hardware mechanisms to boot an enclave (ECREATE, EADD, EXTEND, and EINIT), generate an attestation report and provision secrets (EREPORT), and provide isolated execution (EENTER and EEXIT, and SGX MEE~\cite{intel-mee}).
Similarly, Keystone~\cite{eurosys20-keystone} relies on a security monitor running in the RISC-V machine mode with control over ISA-defined Physical Memory Protection registers in order to enforce memory isolation.
In cloud FPGAs, however, the fabric is spatially shared with the persistent and untrusted Shell logic.
{\it Any and all I/O ports are untrusted}, as applications' custom logic can only connect to a series of I/O ports defined and exposed by the Shell.
}

\noindent\blue{\textbf{Lack of secure and flexible I/O.}
Many FPGA-based accelerators operate on data-intensive problems and stress off-chip I/O with unique security requirements.
For example, some accelerators (e.g., a Bitcoin miner) only require secure register-level access, while other accelerators may exhibit more complex memory accesses including streaming (e.g., for DNN weights) and random-accesses (e.g., for graph processing). 
As we explore in Section~\ref{sec:shield}, each accelerator requires distinct security and performance levels.
Current work on FPGA security~\cite{pfc, eguro-fpl12, cipherbase, meetgo, ambassy, fpga_security_fpga19} does not address the lack of secure I/O as a result of the Shell, nor do they provide security mechanisms that can adapt to the distinct security and performance levels required by diverse accelerators.}

\subsection{Threat Model}

We assume the comprehensive threat model in Figure~\ref{fig:threat-model}. The adversary attempts to compromise the confidentiality and integrity of the code and data processed by the user's accelerator running on the FPGA.
The adversary has {\it full physical control of the FPGA device post-manufacturing} and is {\it able to control privileged CPU software}, such as the OS, VMM, and device drivers.
Furthermore, the adversary is {\it able to control privileged FPGA logic, such as the AWS F1 Shell}.
We assume that {\it any off-chip memory, including HBM, can be compromised}, as the adversary can either perform physical attacks on off-chip buses, or intercept traffic via the Shell logic for HBM.
We assume that the physical package, supply chain, and manufacturer of the FPGA are trusted.
While we use a host program to transfer data, we assume {\it the host CPU is untrusted} and do not depend on any security mechanisms provided by the CPU TEEs.
By removing the host CPU from the TCB, we are not susceptible to the recent attacks that plague CPU TEEs~\cite{meltdown, spectre, sgxpectre, spectre-returns, foreshadow, lvi, zombieload, ridl, fallout, plundervolt, microscope}.

\revised{Side-channel attacks are largely a function of applications' specific logic.
Thus, we do not claim defend against all possible side-channel attacks.
Instead, \SystemName's customizability allows us to provide established tools for developers to mitigate common FPGA side-channel attacks such as controlled channel~\cite{controlled-channel} and power analysis~\cite{zhao-sp18, schellenberg-date18} attacks (see Section~\ref{sec:shield-flexibility}).
We hope the open-source community will leverage \SystemName's flexibility to contribute additional security mechanisms.}

We do not consider denial-of-service (DoS) attacks, as the CSP has physical control of the hardware and  can simply unpower it. The CSP is incentivized to prevent DoS attacks due to the revenue loss when the FPGA instance is unavailable.
We do not consider attacks against the CSP, as the Shell already protects against malicious FPGA users.
\revised{We do not consider covert channels; we assume the tool flow used to generate the accelerator is trusted and run in a secure environment.}

\section{\SystemName Workflow and Security Model}\label{sec:architecture}
\SystemName is an end-to-end framework that enables remote users to design bespoke TEEs for accelerators in cloud FPGAs.
\SystemName is designed around existing mechanisms in cloud FPGAs (i.e., Xilinx UltraScale+ and Intel Stratix 10) and does not require hardware changes.
That said, \SystemName does necessitate the cooperation of both FPGA manufacturers and CSPs to realize the TEE requirements described in Section~\ref{sec:tees-background}.
This section provides an overview of \SystemName and its components, as well as the responsibility of all parties.
Section~\ref{sec:tees} describes how these components enable the requisite TEE building blocks.

The four key parties in the \SystemName framework are shown in Figure~\ref{fig:architecture}.
The \textbf{Manufacturer} is responsible for manufacturing the physical FPGA chip.
The \textbf{Cloud Service Provider (CSP)} owns the physical datacenter that houses servers containing FPGAs and offers them to customers.
\SystemName splits the notion of a ``customer'' into two separate entities.
The \textbf{IP Vendor} creates the actual accelerator design and distributes it to \textbf{Data Owners} (e.g., on a public marketplace~\cite{fpga-marketplace}).
\revised{The Data Owner then rents an FPGA instance from the CSP, programs the accelerator, and uses it to process sensitive data.
The Data Owner should source accelerator designs only from trusted IP Vendors.
Of course, the IP Vendor and Data Owner can be the same entity.
The Data Owner does not have to trust the CSP, but does assume trust in the Manufacturer and the IP Vendor.}

We now review each party's role and how trust is delegated by following the steps in the \revised{\SystemName} framework in Figure~\ref{fig:architecture}.

\textbf{Device Manufacturing (the Manufacturer).}
The security foundation of \SystemName begins with the Manufacturer.
The Manufacturer must provision two keys for each FPGA device: an AES device key and an asymmetric public/private device key pair (e.g., RSA or ECDSA).
The Manufacturer must burn the AES device key into an e-fuse or BBRAM (and optionally encrypt it with a PUF) using existing FPGA security mechanisms \circled{1} during production.
The Manufacturer then embeds the asymmetric private device key into the FPGA SPB firmware and then encrypts the firmware using the AES device key \circled{2}.
The Manufacturer must also register and publish the public device key via a trusted certificate authority.

\textbf{Accelerator Development (the IP Vendor).}
The IP Vendor is trusted to develop the accelerator IP in a secure environment such as a secure workstation
\circled{3}.
During the development process, the IP Vendor connects the accelerator's I/O ports to the open-source \SystemName \emph{Shield} module instead of the Shell interface.
The \emph{Shield} provides configurable runtime secure I/O and isolated execution (see Section~\ref{sec:shield}).
Since the Shield is parameterized RTL logic exposing the same interface as the Shell, the IP Vendor can easily simulate and integrate her design with the Shield.
The IP Vendor can secure multiple accelerator modules with separate Shield modules, enabling multiple isolated execution environments~\cite{amorphos}.

\blue{The IP Vendor then provisions a symmetric Bitstream Encryption Key and an asymmetric Shield Encryption Key for the accelerator.}
The Bitstream Encryption Key is used for bitstream confidentiality, and the Shield Encryption Key is used to protect data transferred between the Data Owner and the FPGA (both explained below).
The IP Vendor embeds the \blue{private} Shield Encryption Key into each respective Shield module and compiles the entire design into a partial bitstream.
Finally, the IP Vendor encrypts the partial bitstream with the Bitstream Encryption Key and distributes the encrypted partial bitstream \circled{4}.
The IP Vendor creates one accelerator bitstream for all users; the attestation process provisions unique keys to each Data Owner.

\textbf{Deployment (the Data Owner).}
Once the Data Owner is ready to process sensitive data using the accelerator, she obtains a cloud FPGA instance from the CSP \circled{5}.
The Data Owner then instructs the CSP's FPGA driver to program the accelerator onto the FPGA.
The FPGA driver first resets the FPGA, which initiates a secure boot process \circled{6}.
The SPB begins by using BootROM code to load the SPB firmware from disk, decrypting it using its embedded AES device key. 
The SPB firmware boots the \SystemName Security Kernel from external storage onto a dedicated Security Kernel Processor executing from its own private, on-chip memory \circled{7}. 
The Security Kernel Processor can either be a reserved hardened CPU in the FPGA or a static bitstream containing a soft CPU (e.g., MicroBlaze or Nios II)~\cite{xilinx-ultrascale-security}.
The SPB firmware hashes the Security Kernel and uses its hash and the private device key to generate a unique asymmetric Attestation Key pair and a corresponding certificate.
As a result, the Attestation Key is cryptographically bound to the FPGA device and the specific Security Kernel binary.
The Security Kernel itself contains no confidential information and has no direct access to device keys, preventing attackers from leaking the device keys via an illegitimate Security Kernel.
The Security Kernel only has access to the Attestation Keys that it receives from the SPB firmware via a secure channel (e.g., on-chip shared memory).

The \SystemName Security Kernel has three primary jobs.
First, it performs remote attestation with the Data Owner and IP Vendor \circled{8} (Section~\ref{sec:remote-attestation}).
Through this process, the Data Owner receives cryptographic proofs showing that the FPGA device, the Security Kernel, and the accelerator partial bitstream are authentic, referencing the Manufacturer and the IP Vendor as certificate authorities, respectively.
Second, it mediates all access to the FPGA fabric.
The CSP uses the Security Kernel to first launch the Shell into its reserved static logic region.
The Security Kernel then securely receives the accelerator's Bitstream Encryption Key from the IP Vendor via a secure channel established during attestation, and decrypts and loads the accelerator onto the FPGA, connecting it to the Shell interface via partial reconfiguration \circled{9}.
Since the Security Kernel is open source and contains no secrets, the CSP can fully control and audit the Shell loading process.
Likewise, the Security Kernel hash is included in the attestation report (Section~\ref{sec:remote-attestation}); the IP Vendor audits the hash before sending over the Bitstream Encryption Key.
Finally, the kernel continuously checks existing hardware monitors.
It can thus detect backdoor activity (e.g., JTAG and programming ports)~\cite{xilinx-ultrascale-security, intel-sdm}, ensure that the authenticated accelerator bitstream is not modified before use, and prevent any physical attacks.
While the Security Kernel relies on the host CPU to communicate with the IP Vendor, this channel is authenticated and encrypted.

\blue{As part of the remote attestation process, the Data Owner generates a symmetric Data Encryption Key for each Shield module (Section~\ref{sec:remote-attestation}).
Data Encryption Keys are used to encrypt the sensitive input data.
The Data Owner receives the IP Vendor's public Shield Encryption Key \circled{10}, and encrypts the Data Encryption Key(s) against the IP Vendor's public Shield Encryption Key to produce Load Key(s).
The Load Key(s) are subsequently used to securely provision the Data Encryption Key into each \SystemName Shield module.
}

Finally, the Data Owner is ready to utilize the accelerator, using an (untrusted) \SystemName host program on the untrusted host CPU to proxy all communication to the accelerator \circled{11}.
The host program forwards the Load Key and the encrypted data to the FPGA.
The \SystemName Shield uses the \blue{private} Shield Encryption Key to decrypt the \blue{Load Key(s) and retrieve the Data Encryption Key(s)}, which in turn secures the user data during runtime. %
When the outputs are ready, the Shield transfers results, encrypted using the Data Encryption Key, back to the Data Owner via the host program.
As is the case with the Security Kernel, all communication through the CPU is encrypted and authenticated.

\section{Securely Enabling TEE Building Blocks}~\label{sec:tees}
We now present a security argument to demonstrate how \SystemName enables all of the TEE building blocks (Section~\ref{sec:tees-background}).

\textbf{Hardware Root-of-Trust.}
While current FPGAs do not provide hardware support for requisite private asymmetric keys, \SystemName is able to build a root-of-trust via two manufacturer-provisioned keys.
The AES device key is the true root-of-trust, protected by existing mission-critical security mechanisms in current FPGAs.
The private device key provides the asymmetric cryptography needed for attestation.
Although it is embedded in firmware, it is encrypted with the AES device key and thus imbued with the same level of trust.

\textbf{Secure Boot.}
\SystemName's secure boot process is built as an extension to current FPGA boot mechanisms, which first executes BootROM code on the SPB.
The BootROM decrypts the SPB firmware using the AES device key and hands off the boot process to it.
We trust that this step is secure as it is deployed in numerous mission-critical FPGA applications.

Once the SPB firmware is initialized, its main job is to bootstrap trust to the Security Kernel running on a dedicated processor.
To do so, it reads the Security Kernel out of the boot medium and hashes it to obtain $H(\texttt{SecKrnl})$.
The SPB firmware signs the hash with the private device Key $\texttt{DeviceKey}_\texttt{priv}$.
It uses the resulting value to seed a key generator to produce a unique asymmetric Attestation Key pair $\texttt{AttestKey}_\texttt{priv,pub}$, which is cryptographically bound to the device and the Security Kernel binary.
It also generates a certificate over the Security Kernel and the resultant Attestation Key by computing $\sigma_\texttt{SecKrnl} = \textrm{Sign}_\texttt{DeviceKey}(H(\texttt{SecKrnl}), \texttt{AttestKey}_\texttt{pub})$.
The SPB firmware then loads the Security Kernel onto the processor and places the Attestation Key pair and $\sigma_\texttt{SecKrnl}$ into the Security Kernel's private memory.
Secure boot occurs before any other software is loaded, ensuring that no untrusted software can tamper with or monitor the Security Kernel.
\blue{If the Security Kernel Processor is a soft CPU, its partial bitstream is hashed alongside the Security Kernel and the bitstream is loaded by the SPB firmware.}

\begin{figure}[t]
  \centering
  \includegraphics[width=3.33in]{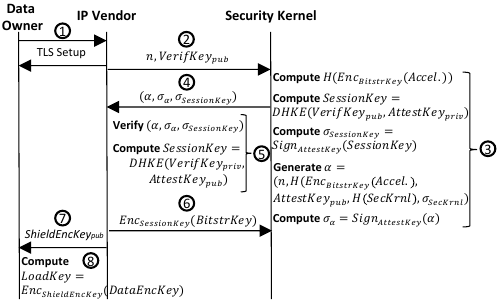}
  \caption{\small Remote Attestation and Secure Storage and I/O Protocols.}
  \label{fig:attest} 
\end{figure}

\textbf{Remote Attestation.}~\label{sec:remote-attestation}
Once the Security Kernel boots, it waits for a remote attestation request.
Remote attestation executes via a series of message exchanges between the Data Owner, the IP Vendor, and the Security Kernel, shown in Figure~\ref{fig:attest}.
Through remote attestation, (a) the Data Owner generates an ephemeral Data Encryption Key used to secure sensitive data, (b) the IP Vendor validates the authenticity of the FPGA device and bitstream, and (c) the Security Kernel receives the Bitstream Key required to load the accelerator. Data Owners are in full control of which IP Vendor to use for remote attestation for each FPGA instance. 

The Data Owner initializes a standard TLS/SSL connection to a trusted IP Vendor server \circled{1}.
The IP Vendor first generates a random nonce $n$, as well as an asymmetric Verification Key pair $\texttt{VerifKey}_\texttt{priv,pub}$.
The IP Vendor sends $n$ and $\texttt{VerifKey}_\texttt{pub}$ to the Security Kernel \circled{2}.

\begin{sloppypar}
Meanwhile \circled{3}, the Security Kernel reads and hashes the appropriate encrypted accelerator bitstream, obtaining $H(\textrm{Enc}_\texttt{BitstrKey}(\texttt{Accelerator}))$.
Using $\texttt{VerifKey}_\texttt{pub}$ and $\texttt{AttestKey}_\texttt{priv}$, the Security Kernel performs key exchange to generate a shared symmetric $\texttt{SessionKey}$ with the IP Vendor, allowing the Security Kernel and the IP Vendor to send encrypted messages.
To prevent any man-in-the-middle attacks, the Security Kernel also signs $\texttt{SessionKey}$ with $\texttt{AttestKey}_\texttt{priv}$ to obtain a new certificate $\sigma_\texttt{SessionKey}$.
The Security Kernel then generates an attestation report $\alpha$, containing $n$, $H(\textrm{Enc}_\texttt{BitstrKey}(\texttt{Accelerator}))$, $\texttt{AttestKey}_\texttt{pub}$, $H(\texttt{SecKrnl})$, and $\sigma_\texttt{SecKrnl}$.
The Security Kernel signs this report using $\texttt{AttestKey}_\texttt{priv}$ to obtain $\sigma_{\alpha}$, and finally sends back $\alpha$, $\sigma_{\alpha}$, and $\sigma_\texttt{SessionKey}$ to the IP Vendor \circled{4}.
\end{sloppypar}

The IP Vendor authenticates the attestation report starting with the $\texttt{DeviceKey}_\texttt{pub}$ received through the Manufacturer's certificate authority \circled{5}.
The IP Vendor checks that $\sigma_\texttt{SecKrnl}$ was signed by the corresponding $\texttt{DeviceKey}_\texttt{priv}$, proving that a legitimate FPGA generated the attestation report. %
To ensure that the Security Kernel (and Security Kernel Processor, if applicable) is valid, the IP Vendor consults a public list of \SystemName Security Kernel (and Security Kernel Processor) hashes.
Next, the IP Vendor authenticates the attestation report by using $\texttt{AttestKey}_\texttt{pub}$ to ensure that $\alpha$ was signed with $\texttt{AttestKey}_\texttt{priv}$.
The IP Vendor matches the signed nonce with $n$, preventing replay attacks, and the signed bitstream hash with $H(\textrm{Enc}_\texttt{BitstrKey}(\texttt{Accelerator}))$, confirming that the correct bitstream was loaded into the Security Kernel's memory.
Finally, the IP Vendor establishes a secure channel to the Security Kernel by first generating the same $\texttt{SessionKey}$ as the Security Kernel using $\texttt{AttestKey}_\texttt{pub}$ and $\texttt{VerifKey}_\texttt{priv}$, and verifying that $\sigma_\texttt{SessionKey}$ was signed by $\texttt{AttestKey}_\texttt{priv}$.

Using the $\texttt{SessionKey}$, the IP Vendor securely transmits the $\texttt{BitstrKey}$ to the Security Kernel \circled{6}.
The Security Kernel decrypts the accelerator bitstream and loads it onto the FPGA, ensuring that the plaintext bitstream containing sensitive IP and Shield Keys are only handled in secure on-chip memory.

\begin{figure*}[t]
\centering
\includegraphics[width=7in]{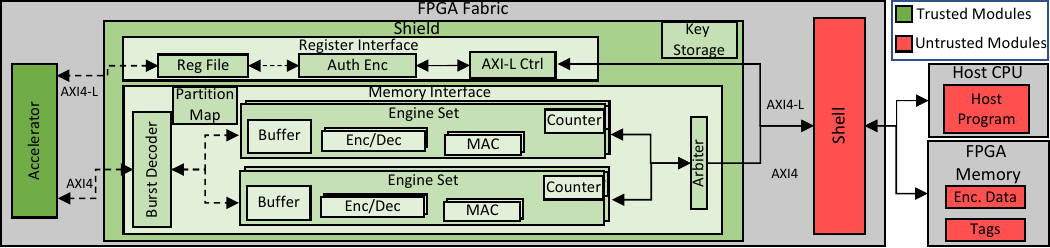}
\caption{\small\SystemName Shield architecture. Dashed and solid lines correspond to unencrypted and encrypted data, respectively.}
\label{fig:shield}
\end{figure*}
\textbf{Secure Storage and I/O.}
\SystemName provides secure storage and I/O to the Data Owner by creating a security perimeter via the \SystemName Shield that encrypts and authenticates all data external to it.
Recall that the IP Vendor provisioned a \blue{private} Shield Encryption Key into each Shield module in the accelerator.
As part of the remote attestation session, the IP Vendor \blue{provides the public Shield Encryption Key to the Data Owner (e.g., via a certificate authority) \circled{7}.} %
\blue{The Data Owner generates at least one Data Encryption Key (e.g., one per Shield module) and encrypts them against the public Shield Encryption Key to get the Load Key(s) \circled{8}.}
The Data Owner then encrypts sensitive input data in a secure location using the \blue{appropriate} Data Encryption Key.
The Load Key(s) are later sent to the FPGA Shield, which decrypts it to get the Data Encryption Key(s).
The Shield uses it to ensure that all sensitive data are secured during both storage and I/O as we discuss in Section~\ref{sec:shield}.

\textbf{Isolated Execution.}
The Shield and Security Kernel ensure isolated execution.
The Security Kernel constantly runs and monitors the FPGA programming and debug ports to prevent tampering with accelerator logic during execution.

\section{\SystemName Shield: Support for Flexible Security}\label{sec:shield}
The Shield is a highly-configurable RTL module that provides isolated execution and secure I/O and storage by interposing on ports between the accelerator and Shell.
The Shield also realizes the necessary \emph{customization} for FPGA TEEs.
IP Vendors configure the Shield to fit their accelerators' memory access patterns and performance and security requirements.

\subsection{Shield Overview}\label{sec:shield-overview}
Figure~\ref{fig:shield} shows how the Shield fits within a typical cloud FPGA deployment consisting of a host CPU and an FPGA accelerator. 
The host program coordinates execution via standard device drivers.
It first loads the Load Key onto the FPGA, which the Shield decrypts into the Data Encryption Key and stores in ephemeral key storage.
The host program then proxies encrypted commands and data between the Data Owner and Shield via the CSP's Shell interfaces.
The host program is untrusted and does not observe any unencrypted data.

The CSP's Shell logic provides two primary interfaces to the host program and the accelerator.
An AXI4-Lite interface, mastered by the Shell, exposes memory-mapped registers for commands and small amounts of data to the host.
The accelerator and host drive an AXI4 and DMA interface, respectively, to access FPGA device memory through the Shell.
\blue{While we focus on securing device memory, Shells commonly provide a generic AXI4 interface for both memory and PCIe~\cite{aws-fpga-dev-kit}.
Thus, the Shield can also support additional interfaces such as PCIe via the same AXI4 interface.}

The Shield provides a wrapper module that transparently secures these interfaces.
The host program accesses registers via AXI4-Lite as before, and the accelerator accesses device memory via the same AXI4 protocol.
The Shield transparently decrypts and encrypts I/O between the host program, the accelerator, and the device memory as shown in Figure~\ref{fig:shield}.
Thus, IP Vendors can simply incorporate and configure the Shield at design time in a plug-and-play manner.
Cryptographic modules that provide authenticated encryption are at the core of the Shield.
We use AES-CTR + HMAC modules as default and present configurable alternatives in Section~\ref{sec:benchmarks}.

{\bf Register interface.}
The register interface provides authenticated encryption using the Data Owner's Data Encryption Key.
The host program memory-maps accelerator-accessible registers and reads/writes encrypted data and commands via pointers.
For writes from the host program, the Shield decrypts and authenticates the data before storing into the accelerator's plaintext register.
Similarly, when the host program reads a mapped address, the corresponding plaintext register is encrypted and tagged before sent to the Shell.
\blue{The Shield provides a register file as shown in Figure~\ref{fig:shield}; users may optionally use their own register file with the Shield simply decrypting/authenticating the AXI4-Lite interface.
Furthermore, because AXI4-Lite register addresses may reveal sensitive information, the Shield offers an additional option of encrypting both addresses and data via a common address for all registers.
In this case, the Shield will decrypt and proxy the data to/from the appropriate plaintext register.
}

\subsection{A Flexible Memory Interface} \label{sec:shield-flexibility}
\subsubsection{Why flexibility is necessary.}
As shown in Figure~\ref{fig:shield}, the Shield exposes the same set of AXI4 interfaces as the Shell.
While the interface is generic, accelerators follow diverse paradigms when using device memory.
Some accelerators stream in large blocks of data from memory, perform random accesses in on-chip buffers, and stream out results to a separate memory region. %
For example, deep neural networks (DNNs) stream in large blocks of weights.
In applications like graph processing, accelerators use non-sequential data-dependent memory accesses.
Even a single accelerator can exhibit distinct paradigms across memory regions, e.g., large streaming reads for weights and smaller reads/writes for activations in a DNN.

The accelerator diversity, in terms of the number of reads/writes to a memory region as well as the amount of data transferred in each burst, have both performance and security implications for the Shield.
First, authenticated encryption, the Shield's core security mechanism, needs to be performed over variable chunk sizes. Smaller chunks require more read requests with higher compute and storage overheads for integrity tags, while overly large chunks transfer unnecessary data bytes. %
It is important to correctly size the chunk granularity for each accelerator's memory regions.

Second, accelerators that read and write the same chunk of memory yield an additional vulnerability.
Computing a MAC over the chunk and its address prevents \textit{spoofing attacks} that directly modify memory contents and \textit{splicing attacks} that copy contents of memory from one address to another~\cite{hardware-memory-authentication}.
However, it does not preclude \textit{replay attacks} where old values are returned for a read, since the corresponding MAC tag is valid.
To safeguard against replay attacks, secure processors resort to Merkle Tree verification schemes~\cite{suh-micro03}, in which MACs are organized in a tree.
Since SRAM is precious in CPUs, only the root node is always kept on-chip, while nodes to a leaf are validated when accessing its corresponding address.

\subsubsection{Designing for flexibility.}
The Shield's memory interface is designed to allow IP Vendors to configure its features and performance, enabling bespoke TEEs customized to each accelerator.
The memory interface consists of one or more \textit{engine sets}.
Each burst request is transformed by a burst decoder in the Shield, which consults a map of IP Vendor-specified memory regions and maps each address range to one of the engine sets.
Each engine set includes encryption and authentication engines alongside on-chip buffers and counters.

The engine set for each memory region can be {\it configured separately} by the IP Vendor to optimize for the different needs and threat models of the target application. We review the configuration options in the current Shield implementation:

\textbf{Cryptographic engines.} Each engine set contains configurable encryption (AES) and authentication (HMAC/PMAC) engines.
The AES engine contains an internal 256-byte lookup table for the S-box.
At design time, the S-box can be duplicated up to 16 times per engine, reducing the AES latency through parallel lookups at the cost of higher resource consumption.
Users are also able to configure the AES key size (128 or 256 bits) during bitstream compilation.
Since the engines expose a simple valid/ready interface, IP Vendors can simply substitute a new cryptographic engine in their place.
We demonstrate this flexibility in Section~\ref{sec:benchmarks} by substituting a PMAC engine in the place of HMAC to enable parallel MAC calculations.
Thus, IP vendors can instantiate multiple instances of each engine to increase parallelism, or even use their own designs.

\textbf{Chunk size.} The IP Vendor denotes a chunk size $C_\text{mem}$ for each memory region, which specifies the granularity of each authenticated encryption chunk.
Each chunk is associated with a 12-byte initialization vector (IV), which is incremented by 1 for each successive chunk to ensure that no two ciphertext blocks reuse the same IV.
Each chunk is authenticated via a 16-byte MAC tag in encrypt-then-MAC mode stored in DRAM.
By using large $C_\text{mem}$ values for streaming patterns, IP Vendors can better amortize the MAC tag overheads.
$C_\text{mem}$ can be any size, from a byte up to the entire FPGA memory.%

\textbf{On-chip buffers.} Each engine set optionally includes a buffer, implemented using Block RAM or UltraRAM, that reduces the overheads for random accesses within small memory regions.
It stores decrypted and authenticated plaintext data and their address ranges, and can be thought as a cache with a line size of $C_\text{mem}$.
If a burst request hits in the buffer, the engine handles all transfers without accessing DRAM.
For misses, the engine set simply generates burst requests to read the entire $C_\text{mem}$-byte chunk and 16-byte MAC tag from DRAM.
The engine set decrypts and authenticates the returned ciphertext in parallel and fill in the buffer line.
If the buffer needs to evict a modified line, the engine set encrypts the line and calculates a MAC tag over the ciphertext, and then performs the necessary writes to DRAM.
For writes, the engine set can first fill the buffer line in the same manner as reads.
Alternatively, if the corresponding chunk is only written to once and not read (e.g., streaming writes), the IP Vendor can simply zero-out the on-chip buffer, avoiding unnecessary reads.

\textbf{Advanced integrity verification.} If an accelerator reads and writes the same chunk multiple times, additional mechanisms are required to prevent replay attacks.
The Shield supports optimized Bonsai Merkle Trees~\cite{bmt} that create {\it Merkle Trees over counters} as opposed to data chunks. 

Merkle Trees are expensive for FPGA designs that need to access \emph{every} tree node from DRAM, unlike CPUs that can benefit from multiple tiers of caches. 
We make the observations that Merkle Trees are used in CPU TEEs because (a) on-chip storage is scarce in CPUs, and (b) secure processors must secure relatively small cacheline-sized chunks (64B) over multiple gigabytes of memory.
Accelerators, however, do not face the same issues, as (a) contemporary FPGAs provide much more on-chip memory via new technologies such as UltraRAM~\cite{ultrascale-products}, and (b) accelerators generally operate on smaller address regions and can leverage large $C_\text{mem}$ chunks.
Thus, \SystemName offers a simpler and more efficient alternative by leveraging the excess of on-chip RAM of FPGAs to store counters only over required address regions (i.e., those that read and write chunks multiple times).

Specifically, an on-chip counter module is configured for these address regions, in addition to the above encrypted authentication mechanisms.
The engine set increments the on-chip counter value $\text{ctr}_i$ by 1 for each write of chunk $i$ to DRAM.
On every read to chunk $i$ from DRAM, a tag $\text{MAC}(i, \text{ciphertext}_i, \text{ctr}_i)$ is generated and verified against the off-chip tag.
In this case, only one extra DRAM access is needed, eliminating excessive off-chip accesses associated with Merkle Trees.
Since the IP Vendor can tailor $C_\text{mem}$, the counter size, and the memory region size to each specific workload, the excess storage overheads are minimized.

\textbf{Side Channels.}
The flexibility of the Shield can also help mitigate \blue{a number of} side-channel attacks discussed in Section~\ref{sec:challenges}.
For controlled-channel attacks, IP vendors can significantly reduce the number of data-dependent memory accesses by increasing $C_\text{mem}$ as an effective countermeasure~\cite{sgx-lapd, controlled-channel}, increasing security by trading off bandwidth and storage efficiency.
\SystemName also provides two effective countermeasures against remote power analysis.
First, \SystemName hides the accelerator's microarchitecture via bitstream encryption, of which current power analysis attacks require significant knowledge~\cite{zhao-sp18, schellenberg-date18}.
Second, \SystemName provides a script to generate an active fence of logic that hides sensitive power signals~\cite{active-fences}.

\blue{However, certain classes of side-channel attacks, such as timing side-channel attacks, are application-dependent and require application knowledge to mitigate~\cite{iodine}.
While we ensure that the timing of Shield cryptographic engines does not depend on any confidential information, we rely on IP Vendors to integrate application-specific techniques if timing noninterference is desired.
We discussed how the register interface can be secured against address metadata attacks in Section~\ref{sec:shield-overview}.
Further security mechanisms against address metadata attacks, such as ORAM~\cite{stefanov-ccs13}, can simply be added by adopting open-source modules (e.g., \cite{tiny-oram}) on top of Shield engines due to their generic interface.
}

\section{Implementation and Evaluation}
\label{sec:eval}
We implemented and evaluated \SystemName on Xilinx UltraScale+ FPGAs.
Since \SystemName only relies on the AES key storage and an SPB, it can also be implemented on Intel FPGAs.
Since we are not allowed to deploy our secure boot process on AWS, we first implemented our end-to-end workflow on a local UltraScale+ Ultra96 FPGA board~\cite{u96}.
We then deployed various Shield configurations with accelerators on AWS F1 instances in order to evaluate performance, assuming correct boot and attestation.

In both cases, we implemented the Shield as portable RTL code in Vivado 2019.2. %
The Shield interfaces with the host program via a \SystemName runtime library, which links against the Xilinx runtime (XRT) that provides FPGA drivers and libraries. %
Thus, as described in Section~\ref{sec:benchmarks}, \SystemName supports accelerators developed using RTL and frameworks such as OpenCL and SDAccel.
XRT and the host program are not in the TCB.

\subsection{End-to-end Ultra96 Implementation}
\label{sec:ultra96-impl}

As mentioned in Section~\ref{sec:architecture}, the root-of-trust is a set of two keys, one AES key embedded in the secure storage and an asymmetric key in the encrypted firmware.
We provisioned an AES key into the Ultra96 e-fuses.
The Ultra96 boot process first executes out of BootROM on the SPB with exclusive access to cryptographic hardware and programming ports.
We embedded the private device key into the firmware and encrypted it with the AES device key.
The firmware is decrypted and loaded onto another hardened processor called the platform management unit.
We boot the Security Kernel on a dedicated Cortex-R5 core running solely from dedicated on-chip memory. %
The Security Kernel communicates with the SPB via a dedicated IPI interface to access cryptographic hardware to generate attestation reports, and runs on the R5 core continuously, monitoring programming and debug ports.

We implemented the full end-to-end \SystemName workflow, using a Bitcoin accelerator (Section~\ref{sec:benchmarks}), on the Ultra96 board. 
We measured that the boot process, from power-on to bitstream loading, completes in 5.1 seconds.
This is relatively small compared to the commonly-observed 40+ second boot time of CSP VM instances~\cite{gce-boot-times, boot-times}, plus the approximate 6.2 seconds of bitstream loading time we observe on F1.
  
\subsection{Shield Evaluation on AWS F1}
\label{sec:shield-eval}
We use AWS F1 instances to evaluate the area and performance overheads of trusted execution with the \SystemName Shield.
The \SystemName Shield introduces encryption and integrity checks to all off-chip memory accesses.
This can create bandwidth bottlenecks if the encryption/authentication rate is lower than the effective data rate of off-chip memory.
This can be addressed by properly configuring the Shield in three ways: a) partitioning the address space (if possible) to use multiple engine sets, b) using multiple AES and PMAC engines within an engine set, or c) increasing AES engine S-box parallelism.
This tradeoff between performance and Shield resource overheads must be carefully managed by the accelerator developer, leveraging the customizable design of \SystemName, in order to achieve the desired security and performance at the lowest cost.

\subsubsection{FPGA Resources Overheads}
\begin{table}
\centering
\caption{\small Shield component utilization on AWS F1. The three base modules on top do not include crypto and on-chip memory (OCM).}
\label{tab:utilization}
\footnotesize
\begin{tabular}{@{}llll@{}}
\toprule
Component                                & BRAM                  & LUT           & REG           \\ \midrule
\multicolumn{1}{l|}{Controller}          & 0 (0\%)               & 2348 (0.26\%) & 547 (0.03\%)  \\
\multicolumn{1}{l|}{Engine Set}          & 2 (0.12\%)            & 1068 (0.12\%) & 2508 (0.14\%) \\
\multicolumn{1}{l|}{Reg. Interface}  & 0 (0\%)               & 3251 (0.36\%) & 1902 (0.11\%) \\ \midrule
\multicolumn{1}{l|}{AES-4x}              & 0 (0\%)               & 2435 (0.27\%) & 2347 (0.13\%) \\
\multicolumn{1}{l|}{AES-16x}             & 0 (0\%)               & 2898 (0.32\%) & 2347 (0.13\%) \\
\multicolumn{1}{l|}{HMAC}                & 0 (0\%)               & 3926 (0.44\%) & 2636 (0.15\%) \\
\multicolumn{1}{l|}{PMAC}                & 0 (0\%)               & 2545 (0.28\%) & 2570 (0.14\%) \\
\multicolumn{1}{l|}{OCM}                 & Variable & 0 (0\%)       & 0 (0\%)       \\ \bottomrule
\end{tabular}
\end{table}
Table~\ref{tab:utilization} shows the FPGA resources (BRAM, LUTs, and registers) used by the Shield components.
\revised{An accelerator's Shield contains one Controller, a configurable number of Engine Sets, and one Register Interface, whose base resource requirements are shown in the top of Table~\ref{tab:utilization}.
The bottom section shows the requirements of various cryptographic engines and on-chip buffers as discussed in Section~\ref{sec:shield}, which are used to augment the Register Interface and Engine Sets.
For encryption, we present two types of AES engines with 4x or 16x parallelism in the S-box (Section~\ref{sec:shield}).
For authentication, we provide a SHA-256 HMAC engine, as well as a PMAC engine based on AES.
Finally, the configurable buffers and counters use a portion of on-chip memory (max available 382Mb).
A full Shield configuration's requirement is a sum of all components and their cryptographic engines and buffers  which combined commonly uses single-digit percents of memory, LUTs, and registers, establishing that \SystemName is cost effective to use with cloud FPGAs.
}

\subsubsection{Throughput Microbenchmarks}
\label{sec:microbenchmark}
\begin{figure}[t]
\centering
\includegraphics[width=3.33in]{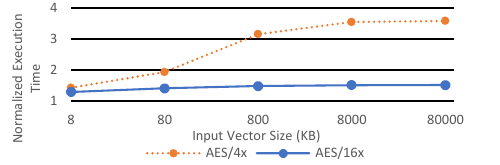}
\caption{\small Vector add throughput overhead for Shield configurations.}
\label{fig:microbenchmark}
\end{figure}
To illustrate the Shield's parallelism, we use a vector-vector add microbenchmark that streams in two vectors and outputs their sum.
The input and output vectors are partitioned and secured with four engine sets each; each set contains one AES-128 and HMAC engine and uses a 512-byte chunk.
The actual logic is minimal and the workload is strictly bound by off-chip memory accesses. 
Figure~\ref{fig:microbenchmark} shows the relative slowdown as a function of the vector size for two Shield configurations.
Slowdown is relative to the execution time without the Shield (insecure version).
For short vectors, execution time is dominated by initialization overheads, e.g., data movement and signalling between the FPGA and CPU.
For long vectors, the overhead directly relates to the encryption throughput available.
By increasing the AES S-box parallelism, the slowdown drops below 50\% for all vector sizes.
We analyzed similarly a matrix multiply microbenchmark, which yielded similar, but less pronounced, insights (maximum overhead of 1.26x for AES/4x) as matrix multiplication involves more computation per data accessed.

\subsubsection{End-to-end Design Example}
\label{sec:sdp}
We now demonstrate how \SystemName can easily enable security for a real-world, end-to-end use case. 
SDP~\cite{vldb21-sdp} recently proposes a GDPR-complaint storage solution by coupling distributed smart Storage Nodes (SNs), to provide encryption-at-rest and line-rate throughput, with a centralized Controller Node (CN), to control access policies and bootstrap SNs.
The authors directly cite the need for an FPGA TEE at each SN.
FPGAs provide both the necessary programmability to adapt to evolving access policies and line-rate throughput difficult to achieve with just CPUs.
Meanwhile, the TEEs protect users' keys and data by a) attesting that the SN is authentic before provisioning keys, and b) encrypting and authenticating application and storage traffic.

SDP thus integrates FPGA TEEs as follows.
On startup, a TEE within each SN remotely attests to the CN.
The CN securely provisions a database of user keys into the TEE.
Applications access files within SNs via TLS, providing a user-specific identity.
Logic within the FPGA TEE on each SN encrypts all traffic to the application (via a TLS key) and storage device (via the user's specific key) for GDPR compliance.

\textbf{Enabling SDP.}
\SystemName directly enables SDP via the workflow presented in Section~\ref{sec:architecture}. %
The GDPR-compliant company assumes the IP Vendor role and creates a bitstream combining a key-value store IP, mapping user identities to files, with the Shield.
SNs securely boot the bitstream and remotely attest their identity to the CN.
The CN encrypts all user keys with the Data Encryption Key and securely provisions them into SN FPGA memory.
Applications then directly form a TLS connection with the Shield, providing a user identity for each file access.
The Shield automatically encrypts and authenticates files between the application and underlying storage. %
By ensuring that each SN is authenticated and securing line-rate I/O traffic, \SystemName allows
the company to deploy high-throughput SNs anywhere in the cloud.

\textbf{Configuring Shield Performance.}
\begin{table}[]
\centering
\caption{\small Performance overhead across Shield designs for SDP.}
\label{tab:sdp-overheads}
\footnotesize
\begin{tabular}{@{}llllll@{}}
\toprule
 \thead{(\# Eng. per Set \\ S-box parall. \\ MAC Module)}
 &
  \thead{4x Eng. \\ 4x\\ HMAC} &
  \thead{4x Eng. \\ 16x\\ HMAC} &
  \thead{4x Eng. \\ 16x\\ PMAC} &
  \thead{8x Eng. \\ 16x\\ PMAC} &
  \thead{16x Eng. \\16x\\ PMAC} \\ \midrule
\thead{\% Overhead} &
  298 &
  297 &
  59 &
  20 &
  20 \\ \bottomrule
\end{tabular}%
\end{table}
While the Shield provides automatic encryption-at-rest, it must be properly configured to meet throughput and security requirements.
To demonstrate this, we created an SDP accelerator that performs gets/puts using a key-value store engine on top of the Shield.
The Shield encrypts and authenticates file accesses via the user key (to storage) and the TLS key (to the application).
Depending on file characteristics, the IP vendor can configure the Shield to authenticate files on a per-file or per-block basis.

We implemented our SDP accelerator on AWS F1 instances and compared Shield overheads to the unsecured key-value store.
Table~\ref{tab:sdp-overheads} shows normalized, steady-state throughput overheads across Shield configurations for 1MB file accesses, using a 4KB authentication block size.
We used two identical engine sets each with 16KB buffer --- one for the storage device and one for TLS.
Instead of using multiple engine sets, which required complex address space partitioning due to variable-length accesses, we increased engine parallelism within each engine set by simply setting a configuration parameter in the Shield.
We began with 4 AES/4x and 1 HMAC engine, but observed high overheads due to the application's high memory intensity.
We attempted to increase S-box parallelism to 16x, but observed limited improvement as HMAC was the bottleneck.
We replaced the HMAC engine with PMAC engines and continued to increase the number of AES/PMAC engines until performance saturated at 8x parallelism for each.
Thus, the Shield's configurability allowed us to enable SDP with only 20\% overhead to line-rate, and with only 4.3\%, 5.0\%, and 2.5\% of the BRAM, LUT, and REG area, respectively.

\subsubsection{Customized Engines for Various Workloads}~\label{sec:benchmarks}
\begin{figure}[t]
\centering
\includegraphics[width=3.33in]{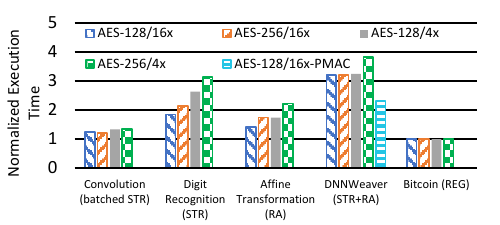}
\caption{\small Execution time of workloads across Shield configurations. Value in parentheses denotes the workload's memory access characteristics (STR=streaming, RA=random accesses, REG=register). PMAC is an additional optimization for DNNWeaver. }
\label{fig:benchmark}
\end{figure}

To demonstrate how the Shield creates a bespoke TEE across accelerators with disparate memory access patterns and security requirements, we use five realistic benchmarks:
a convolutional layer from a neural network with an input size of 27$\times$27$\times$96, a filter size of 5$\times$5, and an output size of 27$\times$27$\times$256 with 32-bit values from a Xilinx reference implementation~\cite{conv};
the digit recognition task from the Rosetta suite using the MNIST dataset~\cite{rosetta};
an affine transformation kernel over 512$\times$512 input images from a Xilinx vision accelerator~\cite{affine};
DNNWeaver using LeNet~\cite{dnnweaver};
and a Bitcoin mining benchmark with a hash difficulty of 24.
To configure the Shield, we instantiated multiple engine sets and engines, depending on the interfaces of each application, and successively increased engine parallelism to saturate memory or accelerator bandwidth.
Figure~\ref{fig:benchmark} shows performance results for four AES engine configurations, with either 4x or 16x S-box parallelism, and 128-bit or 256-bit keys.

\textbf{Convolution} achieves high parallelism by streaming in batches of feature maps and filters, and streaming out each output feature map.
We configure the Shield to match the high parallelism by using 8 engine sets for input images and weights and 4 engines sets for output filters, each with one AES and HMAC engine.
We use a buffer of 128KB in the read set and 64KB in the write set. We configure $C_\text{mem}$ to be 512 bytes to maximize AXI burst lengths for high throughput.
In doing so, we observe small overheads of 1.20x-1.35x.

\textbf{Digit Recognition} also uses separate engine sets for streaming inputs and outputs, but it does not batch.
Hence, we use just 2 engine sets for inputs and 1 engine set for outputs with total 24KB and 12KB buffer, respectively, each with one AES and HMAC engine.
By using a large $C_\text{mem}$ of 512 bytes, we are able to achieve overheads of 1.85-3.15x.
We expect even lower overheads with batching by amortizing input overheads.

\textbf{Affine Transformation} reads non-sequential data, but reads each address once with no writes.
Thus, as with Convolution and Digit Recognition, we can save on-chip memory by disabling integrity counters.
Since Affine Transformation accesses data at consistent chunks of 64B, we use 8 engine sets for inputs with a total 32KB buffer and 4 engine sets for outputs with a total 16KB buffer, all with one AES and HMAC engine.
We observe overheads of 1.41x-2.22x.

\textbf{DNNWeaver} performs both streaming reads for weights and arbitrary accesses for feature maps.
Weights are only read in large chunks, while feature maps require multiple reads and writes for small chunks.
Since DNNWeaver's address space is difficult to partition, we provision two separate engine sets, but increase each set's engine parallelism.
The weights engine set uses a large $C_\text{mem}$ of 4KB, and 4 AES and 1 HMAC engine with total 128KB buffer and no integrity counters.
The feature map engine set uses a smaller $C_\text{mem}$ value of 64B, and similarly 4 AES and 1 HMAC engine with total 64KB of buffer.
As the feature maps cover approximately 1MB of memory, 16KB of on-chip storage is used for integrity counters.
In doing so, we achieve overheads of 3.20x-3.83x. 
These overheads are primarily due to DNNWeaver waiting for long HMAC computations for large 4KB chunks for weights before issuing more bursts.
While the feature map engines also perform integrity checks, we efficiently cache these accesses by configuring a comparatively large buffer.
To alleviate the bottleneck, we replace the HMAC module with 4 PMAC engines in the weight engine set, enabling higher authentication bandwidth for each memory transaction.
This reduces the DNNWeaver overhead to 2.31x for the AES-128/16x case, down from 3.20x.

\textbf{Bitcoin} operates on small data (a 76 byte block header) and only outputs a 4 byte nonce.
We optimize for area by simply leveraging the register interface, with one AES and one HMAC engine, to secure communication.
Because Bitcoin performs significant computation for each input, we observe almost no overheads required to secure the accelerator.

\begin{table}
\centering
\caption{\small Inclusive resource utilization on AWS F1 for the largest Shield configuration across accelerators.}
\label{tab:benchmark-utilization}
\footnotesize
\begin{tabular}{@{}l|lllll@{}}
\toprule
Resource & Convolution & Digit Rec. & Affine & DNNWeaver & Bitcoin \\ \midrule
BRAM     & 2.9\% & 0.71\%     & 2.1\%  & 3.1\%     & 0\%     \\
LUT      & 11\%  & 3.3\%      & 11\%   & 7.1\%     & 1.4\%   \\
REG      & 5.2\% & 1.4\%      & 5.2\%  & 3.5\%     & 0.42\%  \\ \bottomrule
\end{tabular}
\end{table}

Table~\ref{tab:benchmark-utilization} shows the percent resource utilization of the largest \revised{(e.g., AES/16x)} Shield configuration across each accelerator.
By matching the Shield configuration to the patterns and throughput of each accelerator, we provided a TEE with strong security at low performance and area overheads.

\section{Related Work}
\label{sec:related}

{\bf CPU Enclaves.} Numerous flavors of secure processors and CPU enclaves exist~\cite{phantom, sanctum_security16, splittingif_osdi06, oasis_ccs13, isox_micro14, hyperflow-ccs18, xom, suh-micro03, bastion, intel-sgx-explained, arm-trustzone, isca20-park}, with the best known being Intel SGX and ARM TrustZone. Keystone~\cite{eurosys20-keystone} is a recent framework for RISC-V enclaves that addresses CPU-specific memory-management challenges, such as self-paging and dynamic resizing, but does not provide hardware-enabled authenticated encryption. Keystone and other CPU-based enclaves rely on ISA extensions and hardware mechanisms not present in FPGAs, such as RISC-V PMP and SGX instructions. 
\SystemName addresses orthogonal problems. Namely, \SystemName is an end-to-end \textit{framework} securing \textit{arbitrary custom logic} in the face of challenges unique to cloud FPGAs. The flexibility of \SystemName allows users to quickly leverage advances in secure processing, accelerate applications besides CPU enclaves, and quickly address the numerous vulnerabilities arising in secure processors~\cite{sgxpectre, spectre-returns, foreshadow}.

{\bf Accelerator Enclaves.} 
Graviton~\cite{volos-osdi18} modified the GPU peripheral hardware to protect against malicious device drivers from directly accessing sensitive resources, but treats DRAM as trusted memory.
HIX~\cite{jang-asplos19} separated the driver out from the OS and ran it inside a trusted CPU enclave, essentially creating a heterogeneous trusted environment across both CPU and GPU, but requiring changes to the CPU and PCIe root complex. HETEE~\cite{sp20-hetee} proposed fabricating a tamper-resistant box of accelerators (namely GPUs) that a rack of servers can access via a centralized security controller in a dedicated FPGA. However, it did not consider the unique challenges that arise in cloud FPGAs (Section~\ref{sec:challenges}) and required a specialized tamper-resistant chassis.
Telekine~\cite{nsdi20-telekine} addressed a novel side-channel in GPU enclaves in the context of ML training. In comparison, \SystemName assumes a stronger threat model by neither trusting off-chip memory (including HBM) nor relying on a CPU enclave. \SystemName requires no additional hardware or FPGA  modifications. 
\blue{Border Control~\cite{micro15-bordercontrol} sandboxed untrusted accelerators given access to system memory assuming a trusted CPU using similar techniques to \SystemName, such as caching and auditing memory accesses.
\SystemName addresses the opposite problem of protecting accelerators from insecure system software and the Shell.}

{\bf FPGA Security.} FPGAs have been used to accelerate cryptographic primitives~\cite{aes_fpl01, serpent_fpga00}.
There is also increasing interest in general secure computing on FPGAs. 
PFC~\cite{pfc} used proxy re-encryption to provide encrypted I/O to an accelerator, pre-programmed by the manufacturer, on cloud FPGAs.
Cipherbase~\cite{cipherbase} accelerated encrypted database operations using an FPGA.
Eguro and Venkatesan~\cite{eguro-fpl12} used a trusted third party to sign and encrypt bitstreams to be loaded on remote FPGAs.
CPU enclaves have been simulated in FPGAs~\cite{lebedev-csf18,eurosys20-keystone, phantom}.
Commercial FPGAs encrypt bitstreams via an embedded key, preventing adversaries from snooping on or modifying the user design~\cite{xilinx-bitstream-enc}, but requiring a trusted third party.
Coughlin et al. extended this process by using self-provisioning keys in the FPGA hardware to eliminate the third party~\cite{fpga_security_fpga19} and demonstrated a remote attestation protocol.
MeetGo~\cite{meetgo} and AMBASSY~\cite{ambassy} discussed bootstrapping remote attestation to an embedded private key and an ARM TrustZone processor, respectively.
However, these works did not address key challenges described in Section~\ref{sec:challenges}, such as remote attestation within the cloud or the Shell, and required users to implement isolated execution and secure storage and I/O.
Finally, Trimberger et al.~\cite{trimberger-ivsw18} discussed security concerns of cloud FPGAs, including how to detect malicious user logic and prevent tampering with user logic.
Mahmoud et al.~\cite{mahmoud-date19} and Elnaggar et al.~\cite{elnaggar-date19} presented attack methods in multi-tenant FPGAs.

\section{Conclusion}
\label{sec:conclusion}
As compute shifts towards cloud accelerators, the need for both secure and accelerated compute over sensitive data is dire. We address this need with \SystemName, a framework consisting of a secure boot and configurable remote attestation process, as well as Shield logic that guarantees run-time isolated execution. We leverage FPGAs' reconfigurability to allow developers to craft a holistic and bespoke trusted execution environment (TEE) to fit their security and performance requirements. We prototyped \SystemName on current FPGA hardware. We demonstrated secure boot and remote attestation on a local FPGA, enabled a secure storage application, and evaluated the Shield with several accelerators on AWS F1. 

\begin{acks}
We sincerely thank our shepherd, Dmitry Ponomarev, and the anonymous reviewers for their helpful feedback.
We are also grateful to Xilinx for a generous equipment donation.
This work was supported by the Stanford Platform Lab and its affiliates, as well as by a Stanford Graduate Fellowship.
\end{acks}

\appendix
\section{Artifact Appendix}

\subsection{Abstract}
In our artifact, we provide the entirety of the \SystemName source code, including the Shield and implementations of the Secure Boot and Remote Attestation protocols. 
Our artifacts also include a number of reference benchmarks that we use to evaluate \SystemName in Section~\ref{sec:eval}.
We provide instructions on how to build, run, and evaluate Shield benchmarks on AWS F1 instances.
Our archival and GitHub repository also provides a README containing more details on using \SystemName.

\subsection{Artifact check-list (meta-information)}
{\small
\begin{itemize}
  \item {{\bf Algorithm}: \SystemName framework, including Shield, Secure Boot, and Remote Attestation protocols.}
  \item {{\bf Program}: Includes custom vector-vector addition, matrix multiplication, SDP, and Bitcoin benchmarks. Also includes open-source benchmarks from Xilinx~\cite{affine, conv}, Rosetta suite~\cite{rosetta}, and DNNWeaver \cite{dnnweaver}.}
  \item {{\bf Compilation}: AWS F1 benchmarks compiled using AWS FPGA Developer AMI Version 1.8.1, Xilinx Vivado/SDAccel Version 2019.2, and Developer Kit Version 1.4.14. Development and compilation runs on a z1d.2xlarge EC2 instance.}
  \item {{\bf Binary}: Source code for Ultra96 Firmware and Shield included, alongside scripts to generate FPGA bitstreams for AWS F1 instances.}
  \item {{\bf Run-time environment}: AWS F1 bitstreams run on the AWS FPGA Developer AMI Version 1.8.1 with Developer Kit Version 1.4.14, running on an AWS EC2 f1.2xlarge instance.}
  \item {{\bf Hardware}: Shield evaluations performed on an AWS EC2 f1.2xlarge instance.}
  \item {{\bf Execution}: Shield evaluation performed using provided scripts.}
  \item {{\bf Metrics}: Execution latency and resource utilization of various Shield configurations compared to baseline accelerator.}
  \item {{\bf Output}: Execution latency reported by runtime binaries. Resource utilization reported by Xilinx Vivado compilation workflow.}
  \item {{\bf Experiments}: Scripts and detailed workflow provided to compile and generate benchmarks (Figure~\ref{fig:benchmark})}
  \item {{\bf How much disk space required (approximately)}?: About 10 GB for each compiled benchmark.}
  \item {{\bf How much time is needed to prepare the workflow (approximately)}?: Around an hour to set up AWS F1 build environment.}
  \item {{\bf How much time is needed to complete experiments (approximately)}?: Each benchmark bitstream compilation typically takes around 3-4 hours. Compiling all configurations for an accelerator can take a day, although different accelerators may be compiled in parallel using multiple terminals/VMs.}
  \item {{\bf Publicly available}?: Yes}
  \item {{\bf Code licenses (if publicly available)}?: ShEF is held under MIT. Portions (e.g., some source code and benchmark applications) may be held under different licenses, including BSD-2 and Apache License 2.0.}
  \item {{\bf Archived (provide DOI)}?: Yes, at 10.5281/zenodo.5735634 and on GitHub at \url{https://github.com/stanford-mast/ShEF}.}
\end{itemize}
}

\subsection{Description}

\subsubsection{How to access}
\begin{sloppypar}
The source code for the end-to-end ShEF workflow, as well as benchmarks, is archived at 10.5281/zenodo.5735634.
\end{sloppypar}

\subsubsection{Hardware dependencies}
This AE requires access to AWS EC2 instances (z1d.2xlarge and f1.2xlarge).

\begin{sloppypar}
\subsubsection{Software dependencies}
Access to the AWS FPGA Developer AMI (\url{https://aws.amazon.com/marketplace/pp/prodview-gimv3gqbpe57k?ref=cns_1clkPro})
is required, which includes licenses for required Xilinx tools.
Additionally, the AWS F1 Developer Kit is required and can be found on GitHub at \url{https://github.com/aws/aws-fpga}.
\end{sloppypar}

\subsection{Installation}
Obtain the code from the artifact repository or GitHub and follow the instructions in the README under \textit{ShEF Shield Setup}.
Specifically, follow instructions to setup an AWS account and permissions, the Developer instance using a \textit{z1d.2xlarge} instance, and the Runtime instance using a \textit{f1.2xlarge} instance.

\subsection{Experiment workflow}
Specific benchmarks (namely DNNWeaver, bitcoin, and SDP) are located in \texttt{apps/<benchmark\_name>}.
Building and running each benchmark follows a similar workflow.
We describe how to build and run \texttt{dnnweaver\_shield}, a DNNWeaver plus Shield benchmark using LeNet, below.

First, build the bitstream using the Developer instance.
\begin{lstlisting}[language=bash]
# Setup the env variable for the app
cd $SHEF_DIR/apps/dnnweaver_shield
export CL_DIR=$(pwd)

# Setup the build environment
source $AWS_FPGA_REPO_DIR/hdk_setup.sh

# Build the accelerator bitstream
cd $CL_DIR/build/scripts
./aws_build_dcp_from_cl.sh -foreground
\end{lstlisting}

The build script will take several hours (typically 2-3).
Be sure to run in a \texttt{tmux} terminal or omit \texttt{-foreground} to use a nohup context so that SSH session disconnects do not terminate your build.
Once the build is complete, submit your bitstream to AWS.

\begin{lstlisting}[language=bash]
# Use the S3 bucket created during setup
aws s3 cp \
    $CL_DIR/build/checkpoints\
    /to_aws/*.Developer_CL.tar \ 
    s3://<bucket-name>/<dcp-folder-name>/

aws ec2 create-fpga-image \
    --region <region> \
    --input-storage-location \
    Bucket=<bucket-name>,\
    Key=<dcp-folder-name>/<tar-name> \ 
    --logs-storage-location \ 
    Bucket=<bucket-name>,\
    Key=<logs-folder-name>
    
\end{lstlisting}

Save the AFI and AGFI output by the command above.
The AFI build will take around one hour.
Run the following command and check that the status is \texttt{available} before proceeding.
\begin{lstlisting}[language=bash]
aws ec2 describe-fpga-images \
    --fpga-image-ids <your-afi-id>
\end{lstlisting}

Finally, SSH into \textit{Runtime} instance.
Load the FPGA with the AFI, and build and run the test binary.

\begin{lstlisting}[language=bash]
# Assuming same env variables as before
sudo su
source $AWS_FPGA_REPO_DIR/sdk_setup.sh

fpga-clear-local-image -S 0
fpga-load-local-image -S 0 -I \
    <your-agfi-id> # AGFI, not AFI
fpga-describe-local-image -S 0 -R -H

cd $CL_DIR/software
make
./test_lenet
\end{lstlisting}

The README also contains detailed instructions to build and run experiments.

\subsection{Evaluation and expected results}
\begin{sloppypar}
Each benchmark consists of two root application directories, one representing the baseline and one with the Shield.
For example, for DNNWeaver, the projects are at \texttt{apps/dnnweaver} and \texttt{apps/dnnweaver\_shield}, respectively.
Follow the instructions to build and run both as described above (the steps are the same for both applications).
Compare the end-to-end latencies reported by both runtime binaries.
For example, we observe a latency of $5073\mu s$ for \texttt{dnnweaver\_shield} compared to $3054 \mu s$ with \texttt{dnnweaver}.
The resource utilization for a bitstream can be found in \texttt{\$CL\_DIR/build/reports/<timestamp>\ .SH\_CL\_all\_utilization.rpt}, split up by module.
\texttt{shield\_wrapper\_inst} shows the overall Shield resource utilization.
Shield execution latency and resource requirements for a given benchmark are reported in Figure~\ref{fig:benchmark} and Table~\ref{tab:benchmark-utilization}.
\end{sloppypar}

\subsection{Experiment customization}
For each application, the Shield can be customized by modifying parameters included in \texttt{\$SHEF\_DIR/hdk/source/interfaces/\ free\_common\_defines.vh}. 
Please refer to the README for more information.

\subsection{Methodology}

Submission, reviewing and badging methodology:
\begin{sloppypar}
\begin{itemize}
  \item \url{https://www.acm.org/publications/policies/artifact-review-badging}
  \item \url{http://cTuning.org/ae/submission-20201122.html}
  \item \url{http://cTuning.org/ae/reviewing-20201122.html}
\end{itemize}
\end{sloppypar}

\bibliographystyle{ACM-Reference-Format}
\bibliography{main}

\end{document}